# Room-temperature continuous-wave pumped exciton polariton condensation in a perovskite microcavity

Jiepeng Song[1], Sanjib Ghosh[2], Xinyi Deng[1], Qiuyu Shang[1], Xinfeng Liu[3], Yubin Wang[4], Xiaoyue Gao[5], Wenkai Yang[5], Xianjin Wang[5], Qing Zhao[5], Kebin Shi[5], Peng Gao[5], Qihua Xiong[2,4,*], and Qing Zhang[1,*]

[1]School of Materials Science and Engineering, Peking University, Beijing 100871, P.R. China
[2]Beijing Academy of Quantum Information Sciences, Beijing 100094, P.R. China
[3]CAS Key Laboratory of Standardization and Measurement for Nanotechnology, National Center for Nanoscience and Technology, Beijing 100190, P.R. China
[4]State Key Laboratory of Low-Dimensional Quantum Physics and Department of Physics, Tsinghua University, Beijing 100084, P.R. China
[5]School of Physics, Peking University, Beijing 100871, P.R. China
These authors contributed equally: Jiepeng Song and Sanjib Ghosh.
*e-mail: q_zhang@pku.edu.cn; qihua_xiong@tsinghua.edu.cn

**Microcavity exciton polaritons (polaritons) as part-light part-matter quasiparticles, garner significant attention for Bose-Einstein condensation at elevated temperatures. Recently, halide perovskites have emerged as promising room-temperature polaritonic platforms thanks to their large exciton binding energies and superior optical properties. However, currently, inducing room-temperature non-equilibrium polariton condensation in perovskite microcavities requires optical pulsed excitations with high excitation densities. Herein, we demonstrate continuous-wave optically pumped polariton condensation with an exceptionally low threshold of ~0.6 W cm$^{-2}$ and a narrow linewidth of ~1 meV. Polariton condensation is unambiguously demonstrated by characterizing the nonlinear behavior and coherence properties. We also identify a microscopic mechanism involving the potential landscape in the perovskite microcavity, where numerous discretized energy levels arising from the hybridization of adjacent potential minima enhance the polariton relaxation, facilitating polariton condensate formation. Our findings lay the foundation for the next-generation energy-efficient polaritonic devices operating at room temperature.**

Microcavity exciton polaritons (polaritons for short) are bosonic quasiparticles originating from the exciton−photon hybridization in the strong coupling regime, possessing merits of large nonlinearity and small effective mass from their excitonic and photonic parts, respectively[1,2]. They can undergo non-equilibrium Bose−Einstein condensation (polariton condensation for short) through stimulated scattering towards the polariton ground state without the need for population inversion, forming a macroscopically occupied quantum-degenerate phase with spontaneous coherence, which provides an attractive option for developing low-threshold coherent light source[3-12]. For the past two decades, CdTe and GaAs planar quantum well microcavities have served as universal polaritonic platforms for generating polariton condensates[3, 4] and studying their peculiar properties, such as ultralow threshold polariton lasing[13], solitons[14], vortex formation[15, 16], and Kardar−Parisi−Zhang universality[17], albeit normally requiring a liquid helium temperature (~4 K).

Recently, lead halide perovskite semiconductors have emerged as splendid candidates for room-temperature polariton condensation owing to their large exciton binding energy against thermal fluctuation[18-27]. They present unparalleled advantages of solution processability, structural tunability, and high nonlinear interaction strengths compared to other materials with stable excitons at room temperature[28]. Numerous proof-of-principle perovskite polariton condensate devices working at room temperature have been developed, ranging from polariton lasers to all-optical logic elements, XY spin Hamiltonian simulators, and quantum fluids[19, 21-23]. Moreover, perovskite planar microcavities enable the exploration of the rich polariton physics in non-Abelian gauge fields, spin-orbit interactions, and topology, thanks to the remarkable intrinsic material birefringence and the tunable anisotropy in the microcavities[24-27]. On the other hand, while the recent demonstrations of room-temperature polariton condensation under continuous-wave (CW) optically excitation in monolayer transition metal dichalcogenides microcavities have immensely inspired the implementation of ultralow threshold polariton condensate devices[8, 9], the strict fabrication processes and low emission efficiency pose significant challenges to overcome. Emphatically, these issues could be addressed in halide perovskites with excellent emission efficiency and facile fabrication processes. Nonetheless, so far the realization of polariton condensation in perovskite microcavities solely relies on optical pulsed excitations with relatively high energy densities, leaving the question open whether polariton condensation in perovskite microcavities can occur under CW excitation, which can fully exploit the intrinsic low threshold peculiarity of the polariton condensation process.

Here, we report the observation of room-temperature CW optically pumped polariton condensation in a tailored perovskite microcavity, with a low condensation threshold of ~0.6 W cm$^{-2}$ and a narrow linewidth ~1 meV. The observation of macroscopic ground state occupation, superlinear pump density dependence, linewidth narrowing, energy blueshift, linear polarization of the ground state emission, and conclusive evidence of spatial and temporal coherence collectively confirmed the unambiguous transition into the polariton condensate regime. We attribute the remarkably low threshold to the enhanced relaxation process by polariton−polariton scattering, resulting from the interactions among trapped polaritons in the underlying potential landscape. We also experimentally and theoretically investigate the origin of the narrow emission linewidth, ascribing to the hyperfine splitting of discretized energy levels on the ground state. Our results open a route towards developing energy-efficient perovskite coherent light sources and integrated condensate devices in ambient conditions.

Figs. 1a and 1b schematically illustrate the tailored perovskite microcavity under investigation, as well as the coherent emission resulting from the trapped polariton condensates and their interactions. Single-crystalline cesium lead bromide ($CsPbBr_3$) microplatelets were fabricated utilizing an optimized antisolvent method and transferred to the bottom distributed Bragg reflector (DBR) mirror by a dry transfer process using polydimethylsiloxane (PDMS) stamps (see Methods section). A 10-nm-thick polymethyl methacrylate (PMMA) spacer layer was spin-coated to protect the perovskite from direct exposure to air and water; while a thicker layer (~80 nm) would induce a redshift in the cavity mode and an excessive photon component within the polaritons, which weakens the polariton−polariton interaction strength (Supplementary Fig. 1). The microcavity was finally completed by deposition of the top DBR mirror. Remarkably, the PDMS stamp was pressed on the CsPbBr3 microplatelets, causing fractures and the adhesion of surface material, thereby the formation of nanotextured surfaces (Fig. 1c and Supplementary Fig. 2); while the observed pristine samples on the glass substrate possessed atomically smooth surfaces (Supplementary Fig. 3). The separation of nano-ridges ranges from one hundred to hundreds of nanometers and heights of several to tens of nanometers on the sample surface (Fig. 1c). Hence, a trapping potential landscape within the microcavity was established, where the nano-ridges with appropriate dimensions (height ≈ 10 nm, lateral dimension ≈ 300 nm, and separation distance ≈ 300 nm) can result in an enhanced polariton scattering efficiency to achieve polariton condensation (Fig. 1b, details discussed in Fig. 4). Specifically, the interactions among polaritons in adjacent traps result in hyperfine discretized energy levels corresponding to an extended coherence time[29] (~5.1 ps, details discussed in Fig. 3). Meanwhile, they exhibit a broadened distribution in momentum space, lifting the degeneracy for wavevector selection rule during polariton scattering, which enables more efficient scattering towards the ground state and reach polariton condensation[29-31].

The trapping potential landscape was confirmed by the stronger emission from trapped polaritons in certain regions (Fig. 1d and Supplementary Fig. 4), along with the distinct height undulation at the interface between the $CsPbBr_3$ microplatelet and the top DBR mirror (orange dashed outlines in Fig. 1e). Note that the growth byproducts and PDMS residuals can be ruled out during the establishment of the potential landscape (Supplementary Figs. 5-6). Moreover, the buildup of polariton condensation requires microcavity possessing a high quality factor and perovskites of high crystalline quality to guarantee a long polariton lifetime and reduced nonradiative recombination rates, respectively[18]. Owing to the near-unity high reflectivity of DBR mirrors in the stop band (Supplementary Fig. 7), the linewidth of the bare cavity mode reached 1.6 meV at 2.42 eV, indicating a cavity quality factor of ~1512 (Fig. 1f), which is the highest level of the reported perovskite microcavities. Notably, nano-ridges with a height of 10 to 20 nm would not reduce the cavity quality according to the simulations (Supplementary Fig. 8). Fig. 1g shows the room temperature absorption spectrum of the pristine $CsPbBr_3$ microplatelets on the glass substrate (orange), as well as photoluminescence (PL) spectra before (blue) and after (purple) the microcavity assembly. Before being embedded into a microcavity, the microplatelets' absorption spectrum shows a strong inhomogeneously broadened excitonic absorption peak at 2.40 eV, suggesting the presence of stable excitons at room temperature. The corresponding PL emission under non-resonant CW laser excitation is centered at 2.36 eV with a linewidth ($\delta\lambda$) of 68 meV. Meanwhile, an exciton lifetime of 11.7 ns was obtained by the time-resolved PL measurement, implying a superior excitonic quality (Supplementary Fig. 9). Note that no polariton condensation could be observed in microcavities with the microplatelet's exciton lifetime <5 ns, indicating the necessity of the high crystal quality perovskites. After being embedded into a microcavity, two sharp peaks were observed with their linewidths exhibiting considerable narrowing, with $\delta\lambda_1$ = 1.1 meV at 2.272 eV and $\delta\lambda_2$ = 1.0 meV at 2.280 eV, respectively (Supplementary Fig. 10).

Next, we conducted the angle-resolved reflectance and PL spectrum under non-resonant CW excitation of a $CsPbBr_3$ microcavity. The lower polariton branches (LPBs) exhibit clear anti-crossing characteristics, suggesting the occurrence of strong coupling (Fig. 1h and Supplementary Fig. 11). As shown in Fig. 1i, the separation of the spin-split LPB at 0° is ascribed to the *X-Y* splitting arising from the optical birefringence of the orthorhombic-phase $CsPbBr_3$[25] (Supplementary Figs. 12-13). The Rabi splitting energy is fitted as ~128 meV via the coupled harmonic oscillator model (see Methods section), which exceeds the average damping energies of confined photons and excitons, satisfying the strong coupling condition[18]. Moreover, we investigated the angle-resolved PL spectra at room temperature under non-resonant CW (left panel) and femtosecond pulsed laser excitation (right panel) with a comparable pump power of approximately 25 nW (Fig. 1j and Supplementary Fig. 12, see Methods section). The upper polariton branch (UPB) was not observed, which is widely reported in ZnO[32], organic semiconductor[7], and perovskite microcavities[25], likely attributed to the large Rabi splitting, strong continuum absorption, and fast transition from UPB to LPB over the polariton lifetimes[33]. As the room-temperature polariton condensation under pulsed laser excitations was well demonstrated (Supplementary Figs. 14-15), the occurrence of the same phenomenon under CW excitation signified the potential for the sustainment of polariton condensation under ultralow pump power, which would be verified subsequently.

Figs. 2a-c display the angle-resolved PL spectra of the $CsPbBr_3$ microcavity at room temperature, under non-resonant CW excitation at three representative pump densities. At a low pump density of 0.1 W cm$^{-2}$, a broad distribution over the detection angles is observed on the LPB (Fig. 2a). As the pump density increased to 0.6 W cm$^{-2}$, the LPB dispersion at the ground state was occupied massively

with a nonlinear increase of the emission intensity, indicating the onset of polariton condensation (Fig. 2b). As the pump density further increased to 4.5 W cm$^{-2}$, two ground states of the spin-split LPBs dominated in the entire spectrum (Fig. 2c). Meanwhile, Fig. 2d depicts the evolution of the emission spectra measured at 0° with pump densities ranging from 0.1 to 8.0 W cm$^{-2}$. Remarkably, numerous discretized energy levels ($\delta\lambda$ ~1 meV) are distinguished around the ground state even under a rather low pump density, while the spectrum becomes continuous at a high energy region, indicating the influence of the trapping potential landscape on polariton emission in our configuration. Subsequently, as shown in Fig. 2e, the integrated emission intensity of the polariton ground state can be extracted and plotted as a function of pump density in a log-log scale. A clear superlinear increase is observed with the pump density exceeding the condensation threshold ($P_{th}$, ~0.6 W cm$^{-2}$). Moreover, as shown in Fig. 2f, the linewidth narrows from 1.62 to 1.12 meV as the pump density rises across $P_{th}$, suggesting an increased temporal coherence; while it broadens again as the pump density exceeds $P_{th}$ due to decoherence induced by polariton self-interactions[34]. Such narrow linewidth represents almost the highest level among room-temperature polariton condensates (Supplementary Table 1).

On the other hand, an energy blueshift of approximately 0.4 meV is observed at a pump density of 21 W cm$^{-2}$, which extends more slowly to 0.6 meV at 294 W cm$^{-2}$ due to the polariton saturation. The experimental result agrees well with the calculation (orange dashed curve in Fig. 2f, Supplementary Note 1). Meanwhile, this low condensation threshold guaranteed a long-lived, stable, and coherent emission sustained for over 4 hours (Supplementary Fig. 16, see possible solutions for increasing operation time in Supplementary Table 2). In addition, the dependence of $P_{th}$ on the detuning was also investigated, revealing a lower $P_{th}$ as the excitonic fraction increased (Supplementary Fig. 17), which can be attributed to the increased interactions between the polaritons[35]. We also measured the polarization of the polariton condensate emission (Fig. 2g). Two spin-split polariton condensate emissions show mutually orthogonal linear polarization modes with polarization ratios of 0.71 and 0.84, respectively, while the PL emission of the background remains unpolarized (Fig. 2h). The polarization direction is determined by the in-plane optical anisotropy, which already been reported in previous studies[25]. Besides, the pump density-dependent polariton occupation in LPB shows non-equilibrium condensation characteristics (Supplementary Fig. 18). To summarize, the observed macroscopic ground-state occupation, super-linear increase in emission intensity, linewidth narrowing, energy blueshift, and linear polarization of the polariton ground-state emission collectively demonstrate the achievement of exciton polariton condensation under CW pumping at room temperature.

Despite the above manifestations, quantum coherence is another defining criterion of polariton condensation[11]. By utilizing a modified Michelson interferometer, clear interference fringes become distinguishable when the pump density surpasses $P_{th}$ (Figs. 3a-c). The visibility, defined as ($I_{max}$ − $I_{min}$) / ($I_{max}$ + $I_{min}$) with $I_{max(min)}$ the maximum (minimum) intensity of the fringes, reaches a maximum of ~40% at zero time delay, demonstrating a remarkable first-order spatial coherence. Then, the first-order temporal coherence was measured by applying a time delay to the inverted image and extracting the visibility of the central part (see Methods section). Fig. 3d shows the visibility as a function of the time delay at $5P_{th}$. The result can be fitted with a Gaussian function, where a coherent time of ~5.1 ps is extracted corresponding to the polariton condensate emission. The estimated coherence time of the polariton condensate emission is comparable to the values of other reported room-temperature systems[7-9]. Notably, the first-order spatial coherence measurement serves as a convenient and efficient means to promptly ascertain the occurrence of polariton condensation in our microcavities. Moreover, this long-range coherence, combined with the above-mentioned characteristics of polariton condensation, entirely rules out the possibility of random lasing (Supplementary Fig. 19).

Additionally, for the second-order temporal coherence measurement, a classic Hanbury−Brown−Twiss type setup was applied[3, 11], while the ground state of TM spin-split LPB was carefully filtered using an 1800 g/mm grating and tunable bandpass filter to avoid interaction with higher-energy states (Fig. 3a, see Methods section). As shown in Fig. 3e, the measured second-order correlation function at zero time delay, $g^{(2)}(0)$, demonstrated a transition from the thermal ($g^{(2)}(0)$ ~2) to a coherent state ($g^{(2)}(0)$ ~1) with the pump density increase from 0.2 to 2.0 $P_{th}$, providing conclusive evidence of the occurrence of polariton condensation. In this non-resonant CW-pumped configuration, the continuous excitation of the exciton reservoir gradually populates the polariton mode. Consequently, the PL emission below $P_{th}$ primarily comes from a small number of thermal polaritons that shortly relaxed from the excitonic reservoir, which initially would have the same statistical properties as the excitonic reservoir, elucidating the peculiar thermal bunching observed within the nanosecond time scale. Only close to the threshold that the number of polaritons could be enough for a re-thermalization process to develop a new set of thermal properties. Overall, the detected emission well below $P_{th}$ is largely dominated by the properties of the exciton reservoir, which always maintains a large population (compared to the polariton population). Moreover, the ultralow pump density and strong spatial confinement together prevent the convergence of $g^{(2)}(0)$ to a higher value[36]. Remarkably, the slow drop of $g^{(2)}(0)$ towards unity as the pump density surpasses $P_{th}$ could be well reproduced and interpreted by the interactions between trapped polaritons in tightened confinement using a quantum Boltzmann equation (see Methods section, Supplementary Note 2)[37], thus further reflecting the influence of the potential landscape on the spontaneous coherence formation.

Next, we investigate the specific impact of distinct potential landscapes for a better interpretation of the narrow linewidth and ultralow threshold for the room-temperature polariton condensation. Fig.

4a shows the angle-resolved PL spectrum of a CsPbBr$_3$ microcavity with spatial dimensions of 6.6 μm × 13.3 μm in TM polarization (TE polarization and anti-crossing see Supplementary Fig. 20). Two sets of energy levels are observed: one set reveals a separation of approximately 4.0 meV (dashed boxes), attributed to the lateral confinement from the x-direction of the microplatelet edge (~6.6 μm); the other set demonstrates separation of approximately 0.8 meV (arrows), attributed to the interactions among polariton condensates in adjacent traps. Our observation of nano ridges on the perovskite surface that act as photonic potentials triggering this additional lateral confinement, where the observed hyperfine energy levels could be well reproduced by the dissipative Schrodinger's equation based on the two-trap model (Fig. 4b, Supplementary Fig. 21, see Methods section). Although the actual potential landscape is intricate, the successful simulation of experimental observations substantiates the presence of trapping potential landscapes as the primary mechanism driving the generation of the hyperfine discretized energy levels. Additionally, the potential energy profile for the polariton condensate emission is depicted in Fig. 4c, revealing the existence of three discernible groups of discretized energy levels in different spatial regions: 0 < x < 4.2 μm, 4.2 μm < x < 6.8 μm, and 6.8 μm < x < 12.0 μm, respectively. The first and third groups of energy levels originate from two distinct traps, while the middle group is attributed to the interference pattern, where the energy separations are notably smaller and the linewidths are narrower than those in the other two groups (Supplementary Fig. 22). This interference pattern could be also reproduced utilizing our two-trap model (Supplementary Fig. 23, see Methods section).

On the other hand, engineering trapping potentials with spatial distances on the order of the polariton de-Broglie wavelength is challenging for most used spatial light modulators and pattern microfabrication techniques, making the study of polariton condensation in a complex tailored environment still elusive[38]. Here, three typical scenarios with different densities of traps were intentionally fabricated by controlling the pressing force of the PDMS stamp during the dry transfer process, including no traps, few traps, and multiple traps in a similar spatial size, respectively (Figs. 4d-f). As expected, the dispersion of the planar microcavity with no traps has a continuous parabolic shape (Fig. 4g), while quantized states could be observed when traps exist (Fig. 4h), and the hyperfine discretized energy levels occur as the interference pattern originating from the interactions among multiple trapped polariton condensates dominated (Fig. 4i). Moreover, the potential landscape not only affects the emission characteristics of the polariton condensates but also contributes to their ultralow condensation threshold. In this regard, a numerical calculation was carried out to assess the influence of trapping potentials on the polariton relaxation process, where an optimized range of potential landscapes (height ≈ 10 nm, lateral dimension ≈ 300 nm, and separation distance ≈ 300 nm) could yield a great increase in the polariton−polariton scattering rate due to lifting the degeneracy for wavevector selection rule[29-31], resulting in the low condensation threshold (Supplementary Note 3, Supplementary Figs. 24 and 26).

In conclusion, we demonstrated CW optically pumped polariton condensation at room temperature in tailored perovskite microcavities assisted by the potential landscape engineering. The condensation threshold reaches a record-low magnitude of nearly 0.6 W cm$^{-2}$ and the emission linewidth narrows down to as small as 1 meV. The interactions among trapped polaritons in the potential landscape reinforced the relaxation process and thus resulted in an ultralow condensation threshold. Meanwhile, the interactions between trapped polariton condensates in the potential landscape resulted in hyperfine discretized energy levels, arising from the hybridization of the lowest energy mode of the adjacent traps. Our straightforward and practical potential landscape engineering offers an ideal testbed for low-threshold coherent light sources and energy-efficient integrated polaritonic devices operating at room temperature.

## Methods
### Materials preparation
The CsPbBr$_3$ microplatelets were synthesized using an optimized antisolvent method. Analytically pure CsBr and PbBr$_2$ powders with a molar ratio of 1:1 were mixed and dissolved into dimethyl formamide solvent (concentration: 20 mmol L$^{-1}$) as precursors. Glass substrates were placed in a culture dish and nested in a beaker containing dichloromethane as the antisolvent. The precursor solution was then deposited onto the glass substrate, covering it completely. Subsequently, to control the evaporation rate, the beaker was sealed by tinfoil, with an optimized number of uniformly distributed holes on the top sealing. After that, the reaction was maintained at a steady temperature of 35°C for over 24 hours, resulting in the formation of CsPbBr$_3$ microplatelets.

### Microcavity fabrication
The bottom DBR mirror was fabricated by thermal evaporation deposition, involving 19.5 pairs of titanium oxide and silicon dioxide. A 154-nm-thick silicon dioxide layer was subsequently deposited as the spacer, followed by transferring CsPbBr$_3$ microplatelets onto the bottom DBR mirror through a dry-transfer method using PDMS stamps. A 10-nm-thick PMMA spacer layer was then spin-coated on the CsPbBr$_3$ microplatelets surface with a spin speed of 4000 rpm. The microcavity was finally completed with the deposition of the top DBR mirror, which consisted of 12.5 pairs of silicon dioxide and tantalum pentoxide.

### Optical spectroscopy characterizations
A custom-built setup utilizing a Fourier imaging configuration was employed to enable real-space and momentum-space resolved spectroscopy and imaging. The emission from the perovskite microcavity was collected using a 50× objective (NA = 0.80) or 100× objective (NA = 0.90), then sent to a spectrometer with a Peltier-cooled charge-coupled device camera (Horiba iHR320). The excitation source was a 405 nm non-resonant CW laser with a pump spot size of 2 μm. For optical pulse excitation, a 400 nm pulsed laser (pulse duration: 100 fs, repetition rate: 1 kHz) generated by the double frequency of an 800 nm laser from a Coherent Astrella amplifier through a beta barium borate crystal was used. For the first-order spatial and temporal coherence measurements, a Michelson interferometer was coupled to the detection path. The emission pattern was divided into two interferometric arms, where one arm installed a prism for horizontal coordinate inversion, and the other was mounted on a nanometric translation stage to control the temporal delay accurately. For the second-order coherence measurement, a Hanbury-Brown-Twiss type setup was coupled to the detection path, including two silicon single-photon avalanche diodes (Micro Photon Devices, $PD-100-CPC), one time-correlated single-photon counter (PicoQuant Hydraharp 400), and one tunable bandpass filters (Semrock, TBP01-561/14-25×36). For time-resolved PL measurements, a frequency-doubled 400 nm pulsed laser (pulse duration: 120 fs, repetition rate: 80 MHz) from Coherent Mira 900 was used as the excitation source, and the time-correlated single-photon counter above mentioned was used to analyze the signal. All experiments were conducted at room temperature.

**Other characterizations**
AFM images are obtained in the tapping mode using a BRUKER Dimension ICON. For the laser confocal microscope image, an Olympus microscope system (FV3000) is used. The cross-sectional samples are prepared using the Helios G4 UX DualBeam™ with 30 kV Ga$^+$ ions. High-resolution cross-sectional HAADF-STEM and energy-dispersive X-ray (EDX) elemental mapping images are obtained using an aberration-corrected scanning transmission electron microscope (FEI Titan Cubed Themis G2 300, operated at 300 kV acceleration voltage).

**Theoretical model**
Our theoretical description of microcavity polaritons is given by the driven-dissipative Schrodinger's equation,

$$i\hbar \frac{\partial \psi_S}{\partial t} = \left[ \Delta_S - \frac{\hbar^2}{2m_{xS}} \frac{\partial^2}{\partial x^2} - \frac{\hbar^2}{2m_{yS}} \frac{\partial^2}{\partial y^2} + V(r) + \frac{i(\gamma - P)}{2} \right] \psi_S + \frac{\Omega}{2} \varphi$$

$$i\hbar \frac{\partial \varphi}{\partial t} = E_{ex} \varphi + \frac{\Omega}{2} \psi_S$$

These coupled light–matter equations represent the emergence of polaritons from the strongly coupled cavity photons and excitons in our perovskite microcavity at room temperature. The wavefunction $\psi_S$ represents the cavity photons with linear polarization $S$ (which could be either X or Y), while $\varphi$ represents the exciton wavefunction. The parameters $\Delta_S, m_{xS}, m_{yS}, \gamma, P, E_{ex}$ and $\Omega$ represents linear spin splitting, photon masses along the x and y axes, photon decay rate, incoherent continuous-wave pump, exciton energy and Rabi splitting respectively. The eigenvalues and the eigenvectors of the non-Hermitian Hamiltonian of the strongly coupled cavity photons and excitons are used to calculate polariton dispersions, which are calculated using a two-trap potential $V(r)$.

The quantum statistics of polaritons can be described by a quantum Boltzmann equation:

$$\frac{d}{dt} P_{n_1 n_2 \ldots n_N} = \gamma \sum_j (P_{n_1 n_2 \ldots n_j - 1 \ldots n_N} - P_{n_1 n_2, n_3 \ldots n_N}) + \sum_j \eta_j (P_{n_1 n_2 \ldots n_j - 1 \ldots n_N} - P_{n_1 n_2, n_3 \ldots n_N})$$
$$+ \sum_{ij} W_{i \to j} n_i (n_j + 1) P_{n_1 n_2 \ldots n_N}$$

where $\gamma$ is the constant decay rate of the quantum states, $\eta_j$ are the effective gains of the quantum modes and $W_{i \to j}$ are the scattering rates. Phenomenologically, we find that low energy quantum modes and their interactions through the scattering rates $W_{i \to j}$ can be effectively described with effective gain rates $\eta_j$ which are proportional to the applied pump fluence $P$, and a constant scattering rate for all the modes. A constant effective scattering rate can originate from the strong disorder which breaks the translation invariance and relaxes the momentum selection rules. The zero-time second-order correlation function is given by,

$$g^{(2)}(0) = \frac{\langle n_1^2 \rangle - \langle n_1 \rangle}{\langle n_1 \rangle^2}$$

where the expected values are defined by, $\langle O \rangle = \sum_{n_1, n_2 \ldots n_N} O(n_1, n_2, \ldots, n_N) P_{n_1 n_2 \ldots n_N}$. The considered parameters are $\Delta_{X/Y} = \pm 4$ meV, $m_{xS} \approx 5 \times 10^{-4} m_e$, $\frac{m_{xY}}{m_{yX}} \sim 0.71$, $W_{i \to j} = 2$ meV, $E_{ex} \approx 2.47$ eV, and $\Omega = 120$ meV, where $m_e$ is the electron mass. We considered $N = 10$ low energy quantum modes in the discretized polariton dispersion originating from the local trapping potentials.

**Data availability**
Data that support the findings of this study are available from the corresponding author upon reasonable request.

**Code availability**
The custom code employed in this work to perform all calculations is available from the corresponding author upon reasonable request.

**Acknowledgments** Q. Zhang acknowledges the support of the National Natural Science Foundation of China (52072006, 51991340, and 51991344), Beijing Natural Science Foundation (JQ21004), and the Ministry of Science and Technology (2017YFA0205700, 2017YFA0304600, and 2017YFA0205004).

**Author contributions** Q. Zhang and J. S. conceived the project. S. G. performed the theoretical calculations. J. S. and Q. S. prepared the perovskite samples. X. D, Q. S., and J. S. fabricated the microcavity. J. S. performed the angle-resolved spectroscopy and imaging, first-order spatial and temporal coherence measurements, time-resolved PL measurements, AFM characterizations, and other general characterizations. Y. W. and J. S. conducted the second-order temporal coherence measurements. X. G., J. S., and P. G. conducted the HAADF-STEM and STEM-EDX elemental mapping. W. Y., X. W., J. S., Q. Zhao, and K. S. conducted laser confocal microscopy imaging. J. S., S. G., Q. S., X. L., Q. X., and Q. Zhang analyzed the data. J. S., S. G., X. D., Q. X., and Q. Zhang wrote the manuscript. Q. Zhang supervised the whole project. All the authors discussed the results and revised the manuscript.

**Competing interests** The authors declare no competing interests.

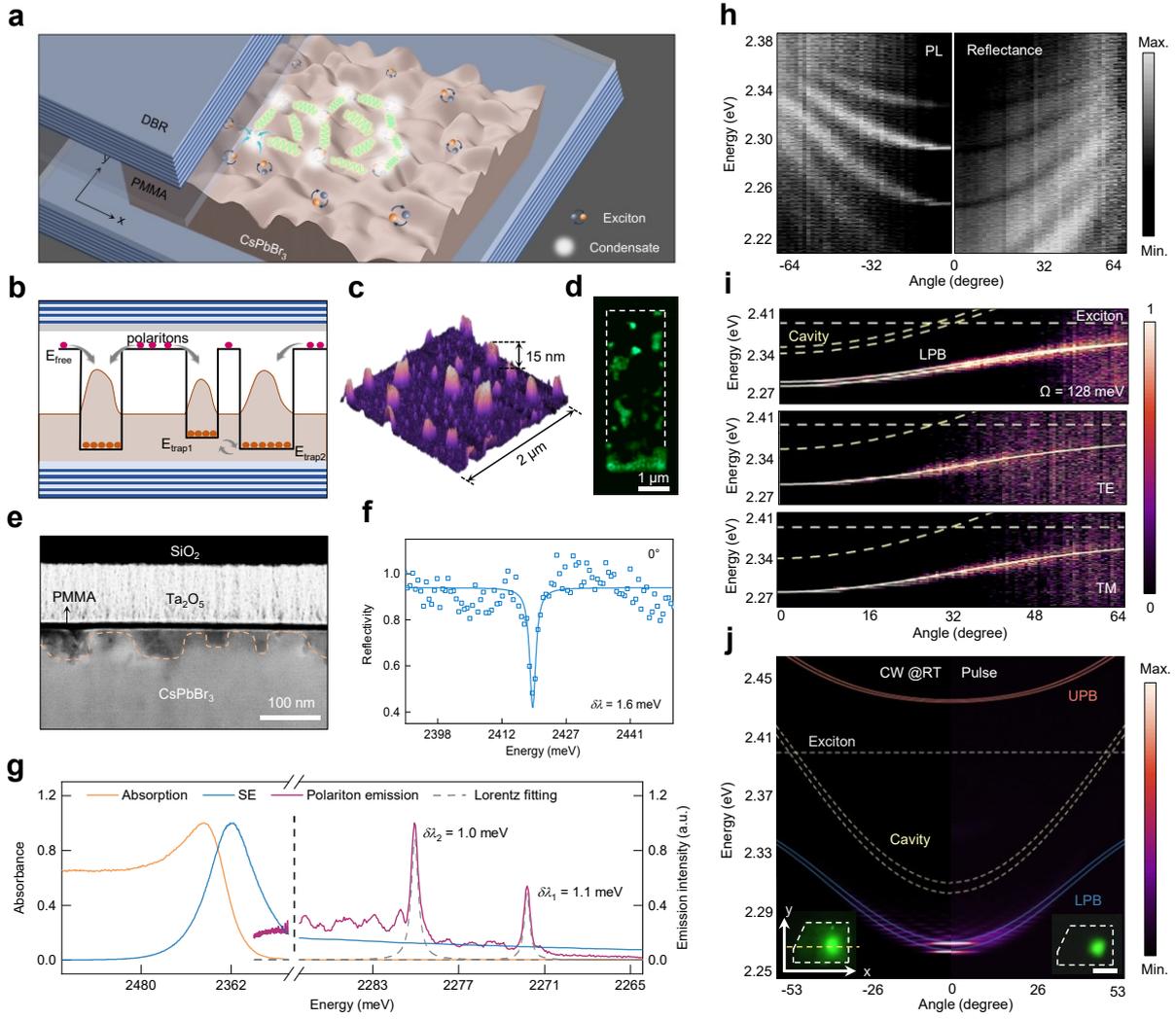

**Fig. 1 | High-quality tailored perovskite microcavity. a-b,** Schematic of trapped polariton condensates and their interactions in a nanotextured perovskite microcavity. **c,** AFM 3D image of the stochastically nanotextured surface of a CsPbBr$_3$ microplatelet before the microcavity assembly. **d,** Laser confocal microscopy image of the perovskite microcavity (excitation source: 405 nm CW laser, collected spectral range: 530 to 550 nm). **e,** High-resolution cross-sectional high-angle annular dark-field scanning transmission electron microscopy (HAADF-STEM) image of the interface between the CsPbBr$_3$ microplatelet surface and top DBR mirror. **f,** Reflectivity spectrum of the bare microcavity without perovskite inside at normal incidence ($\theta = 0°$). The linewidth $\delta\lambda$ is 1.6 meV at 2.42 eV, corresponding to a quality factor of ~1512. **g,** Room-temperature PL spectra of a CsPbBr$_3$ microplatelet on the glass substrate (blue, $\delta\lambda = 68$ meV at 2.36 eV) and embedded into the microcavity (purple, two sharp peaks with $\delta\lambda_1 = 1.1$ meV at 2.272 eV and $\delta\lambda_2 = 1.0$ meV at 2.280 eV). The orange curve is the absorption spectrum of the CsPbBr$_3$ microplatelets on the glass substrate, showing a strong excitonic peak at 2.40 eV. **h,** Angle-resolved PL and the corresponding reflectance spectra of a CsPbBr$_3$ microcavity, showcasing anti-crossing characteristics. Note that the high cavity quality factor (>1500) may cause the vagueness of the reflectance spectra because of the reduced signal intensity. **i,** Angle- and polarization-resolved PL spectra of a CsPbBr$_3$ microcavity at room temperature, under non-resonant CW excitation with the pump density slightly above $P_{th}$. Lower polariton modes in TE and TM polarizations are fitted based on the coupled harmonic oscillator model (middle and bottom panel). White and yellow dashed curves denote the uncoupled exciton and bare cavity mode. A Rabi splitting energy of ~128 meV can be extracted. **j,** Angle-resolved PL spectra of an identical CsPbBr$_3$ microcavity at pump density of ~1.5 $P_{th}$, excited by the non-resonant CW (left) and femtosecond pulsed laser (right), respectively. Inset: the corresponding real-space PL images. Scale bar: 4 μm.

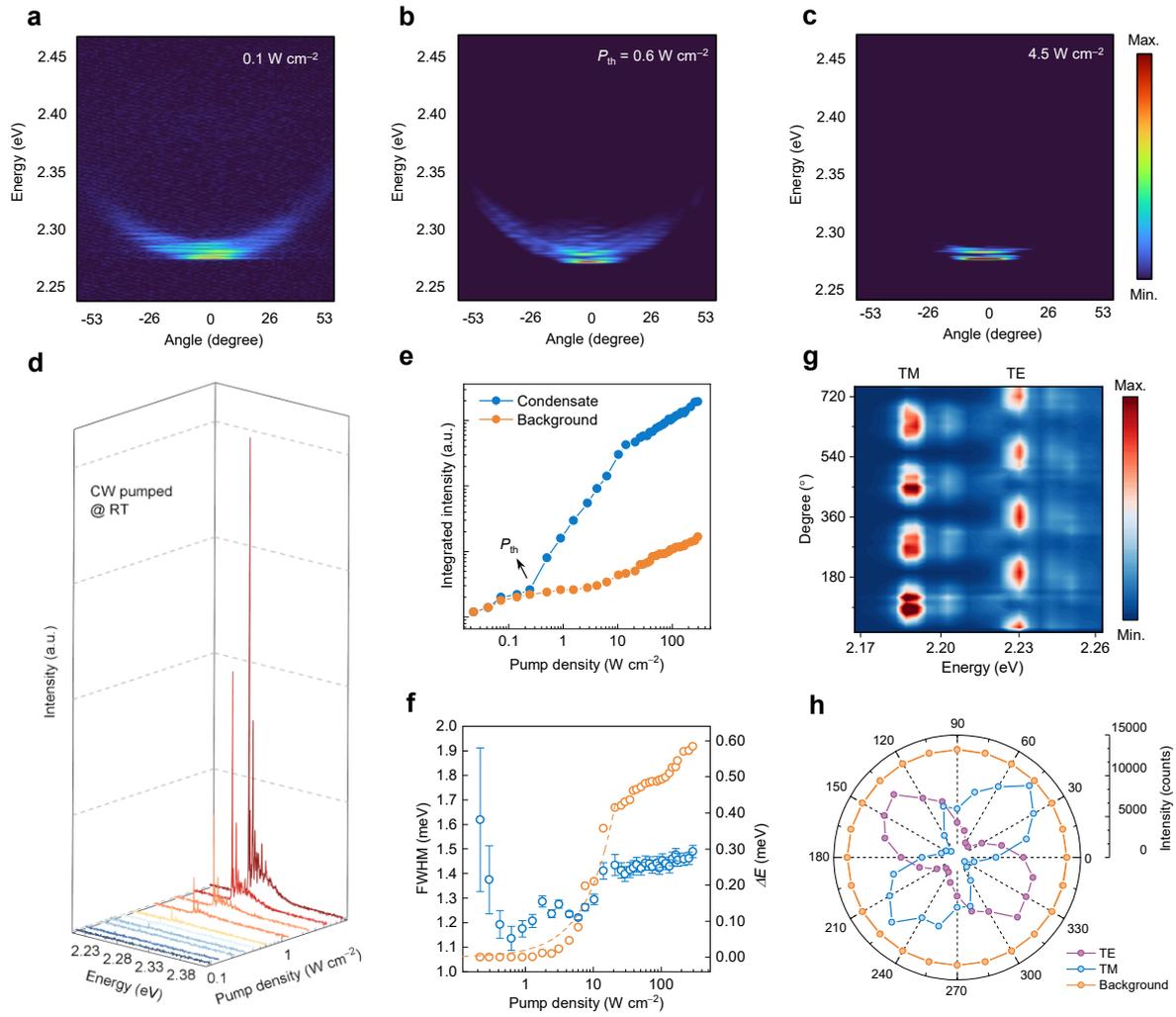

**Fig. 2 | CW pumped polariton condensation at room temperature. a-c,** Angle-resolved PL spectra under non-resonant CW excitation at pump densities of $0.17P_{th}$ (**a**), $P_{th}$ (**b**), and $7.5P_{th}$ (**c**), respectively. **d,** Evolution of emission spectra measured at 0° with the pump density ranging from 0.1 to 8.0 W cm$^{-2}$. **e,** Integrated emission intensity in the condensate energy interval as a function of the pump density in a log-log scale (blue), and also the background PL emission (orange) for reference. **f,** Linewidth (blue hollow dots) and energy blueshift relative to the polariton emission energy at the lowest pump density (orange hollow dots) as functions of the pump density. The orange dashed curve is the calculation result of the blueshift. **g-h,** Above $P_{th}$ two-dimensional pseudo-color polarization plot of the polariton condensate emission at 0° (**g**) and the corresponding polar plot (**h**), suggesting two mutually orthogonal linear polarization modes (purple: transverse-electric mode, TE; blue: transverse-magnetic mode, TM), while the background emission remains completely unpolarized (orange).

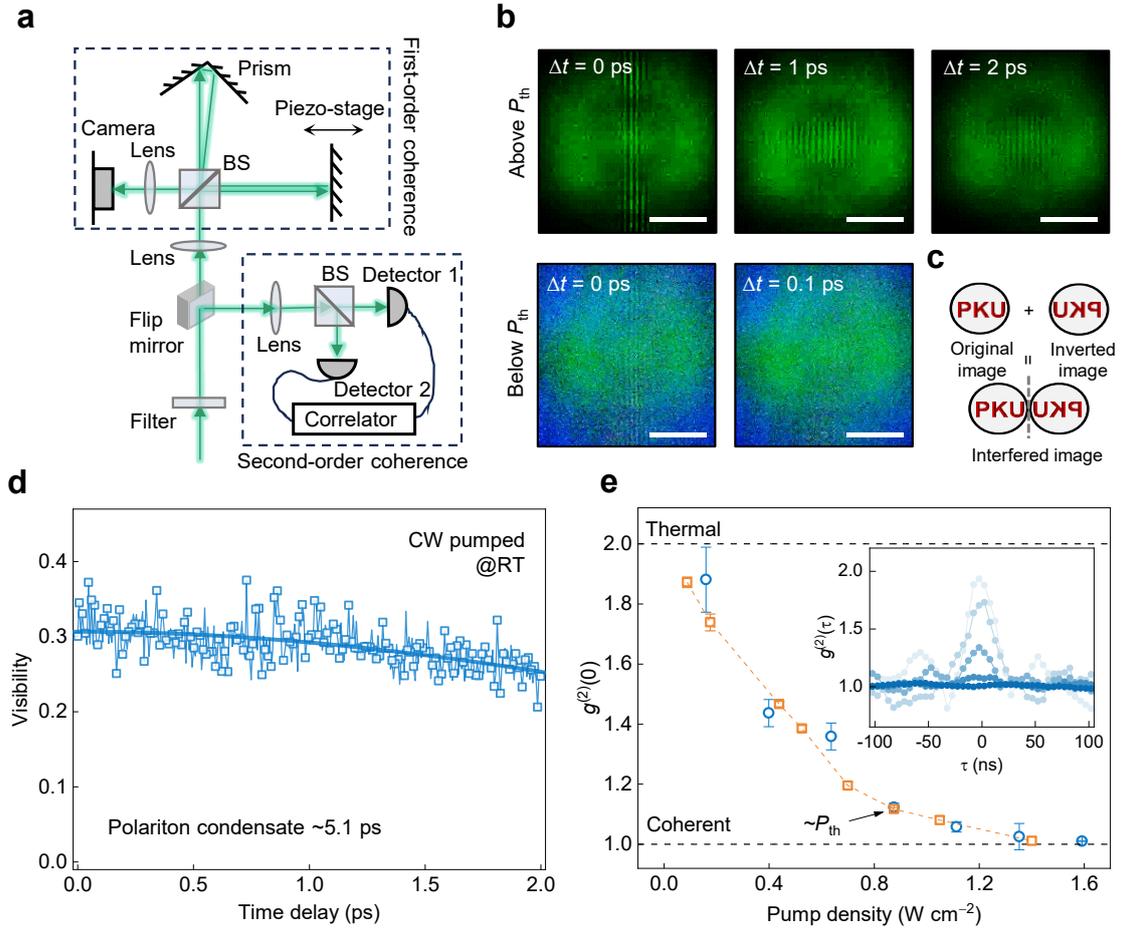

**Fig. 3 | Conclusive evidence of spatial and temporal coherence. a,** Experiment setup sketch of the coherence measurements. BSs are 50:50 beam-splitters. **b,** Michelson interference patterns under non-resonant CW excitation at room temperature. Top panel: above $P_{th}$ with a time delay ($\Delta t$) of 0, 1, and 2 ps, respectively. Bottom panel: below $P_{th}$ with $\Delta t$ of 0 and 0.1 ps, respectively. **c,** Illustration of the mirror-symmetrically interfered image, which is a composite of the original image and its inverted counterpart. **d,** Visibility of the interference fringes as a function of the time delay above $P_{th}$. A coherence time of polariton condensate emission of ~5.1 ps can be extracted. **e,** $g^{(2)}(0)$ as a function of the pump density, demonstrating a transition from the thermal state to the coherent state. Blue and orange hollow dots denote experimental data and calculated results, respectively. Inset: evolution of experimental $g^{(2)}(\tau)$ with pump densities increasing from 0.1 W cm$^{-2}$ (light cobalt blue) to 1.6 W cm$^{-2}$ (dark cobalt blue).

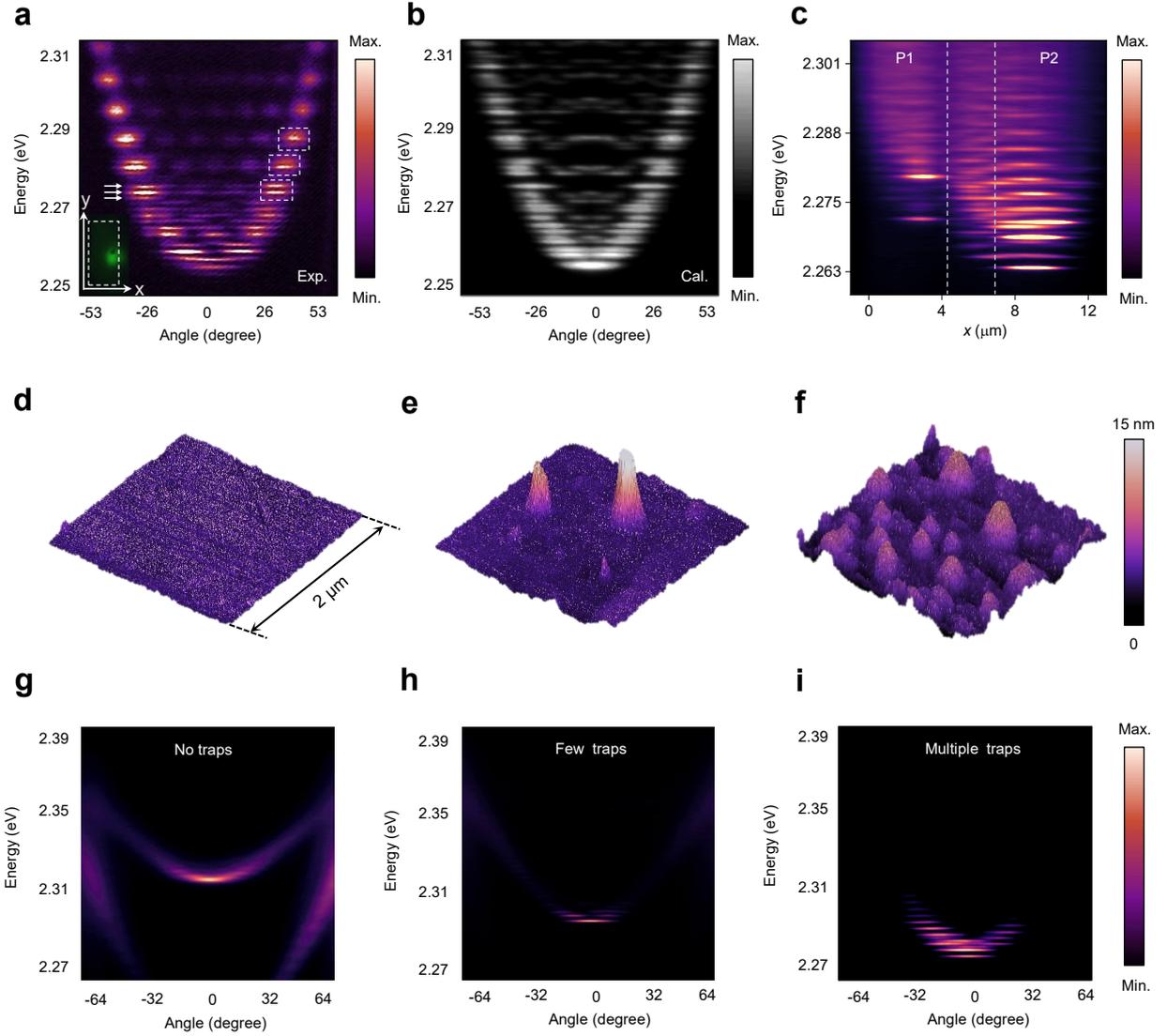

**Fig. 4 | Polariton condensate manipulation by trapping potential engineering. a,** Angle-resolved PL spectrum in TM polarization of a CsPbBr$_3$ microcavity when the short axis x (~6.6 μm) is set to be parallel to the entrance slit of the spectrometer. Dashed boxes: three typical quantized states with separation of ~4.0 meV, deriving from the lateral confinement of the edge of the microplatelet along the x-direction. Arrows: three typical hyperfine peaks with separation of ~0.8 meV, deriving from the interactions among polariton condensates in adjacent traps. Inset: the corresponding real-space PL image. **b,** Calculated angle-resolved PL spectrum based on the two-trap model. **c,** Spatially-resolved PL spectrum at 0° along a horizontal line (yellow dashed) in the inset of Fig. 1j. Two groups of discretized peaks (P1 and P2) with an energy difference of ~9 meV are recognized. Hyperfine states between them are attributed to the interactions between two adjacent trapped polariton condensates. **d-f,** AFM 3D images of the CsPbBr$_3$ microplatelet (spatial dimensions >20 μm) surfaces before the microcavity assembly, with no traps (**d**), few traps (**e**), and multiple traps (**f**), respectively. **g-i,** The corresponding angle-resolved PL spectra of (**d**)-(**f**) in TM polarization under non-resonant CW excitation, with the pump density slightly above $P_{th}$ for (**h**) and (**i**).